# Effect of alloying additions on the lattice ordering of Ti$_2$AlNb intermetallic

Adilakshmi Chirumamilla and Gopalakrishnan Sai Gautam*

Department of Materials Engineering, Indian Institute of Science, Bengaluru 560012

*Corresponding author: saigautamg@iisc.ac.in

## Abstract

Alloys based on the orthorhombic-Ti$_2$AlNb intermetallic phase (O-phase) are promising materials for high-temperature applications in jet engines, given that they can potentially replace Ni-based superalloys in some operating regions of the engines. However, the O-phase is prone to lattice disordering at high temperatures, primarily via anti-site defect formation across the Ti and Nb sites, which can reduce the material's creep resistance and high-temperature tensile properties, necessitating the need to identify strategies to mitigate the disorder. Here, we focus on identifying suitable alloying additions to suppress the disordering of the O-phase using density functional theory and nudged elastic band calculations. Specifically, we consider six different alloying additions, namely, V, Cr, Fe, Mo, Ta, and W, and examine their role in the thermodynamics of anti-site formation and the kinetics of atomic diffusion. Upon verifying the ground state structure and formation energy of Ti2AlNb, we observe the proclivity of all alloying elements (except V) to occupy the Nb site in the O-phase structure. Subsequently, we find that none of the alloying additions can effectively suppress anti-site formation in Ti$_2$AlNb, highlighting the unfavourable thermodynamics. However, we find that Mo and W additions to Ti$_2$AlNb can kinetically suppress the disorder by reducing the diffusivities of Ti and Nb, by ≈4× and 8× compared to the pristine O-phase, respectively, at an operating temperature of 823 K. Thus, Mo and W additions represent a promising strategy to improve the creep resistance of Ti$_2$AlNb-based alloys.

# 1 Introduction

Ti alloys are favored in aerospace applications as structural components due to their exceptional specific strength and resistance to oxidation and corrosion. In the compressor region of commercial jet engines, temperatures can reach up to 700°C. Conventional Ti alloys can only withstand temperatures up to 550°C, beyond which they undergo environmental embrittlement caused by exposure to oxygen at elevated temperatures,[1] thus limiting their usability. Consequently, conventional Ti alloys are utilized in the 'front' stages of jet engines where the temperatures are lower, while Ni-based superalloys are employed in stages where temperatures exceed 500°C.[2] However, the utilization of Ni-based superalloys adds considerable weight to the components. Alternatively, alloys based on titanium aluminides, specifically γ-TiAl and $Ti_3Al$ intermetallic compounds, exhibit enhanced resistance to creep and excellent oxidation resistance, potentially making these phases useful in the back stages of jet engines as well. However, such Ti-Al alloys have been limited by their reduced ductility and formability at room temperature. Thus, mitigating high temperature creep while not sacrificing the room temperature formability remains an ongoing challenge in the development of Ti-Al alloys.

In attempts to improve the room-temperature ductility and fracture toughness of the $Ti_3Al$ intermetallic, Nb has been identified as a useful addition, yielding the $Ti_2AlNb$ intermetallic phase. The ternary $Ti_2AlNb$ phase is known as the O-phase, owing to its orthorhombic structure with *Cmcm* symmetry.[3] To describe the O-phase, a three sub-lattice model is used with α, β, γ sublattices representing 8g (Ti, blue circles), $4c_2$ (Nb, green), and $4c_1$ (Al, orange) sites, respectively, as shown in **Figure 1**a. The O-phase is stable up to 1000°C, and alloys based on the O-phase show better room-temperature ductility and fracture toughness than conventional titanium aluminides with negligible loss in high-temperature properties.[4]

The O-phase exhibits two distinct variants: $O_2$ (ordered) and $O_1$ (disordered), which are characterized by identical space group and lattice periodicity but differing site occupations across the three sub-lattices.[5] In the ordered state (**Figure 1**a), the 8g and the $4c_2$ sites are occupied by Ti and Nb atoms, respectively, while in the disordered state (**Figure 1**b), random occupancy of Ti and Nb atoms is observed at both the 8g and 4c$_2$ sites (purple circles in **Figure 1**b). In both variants, the 4c$_1$ sites are fully occupied by Al atoms. Thus, the O$_2$ phase is susceptible to disordering within the α (8g) and β (4c$_2$) sublattices, involving the exchange of Ti and Nb atoms, leading to the formation of anti-site defects that can eventually result in the O$_1$ phase.[5-9]

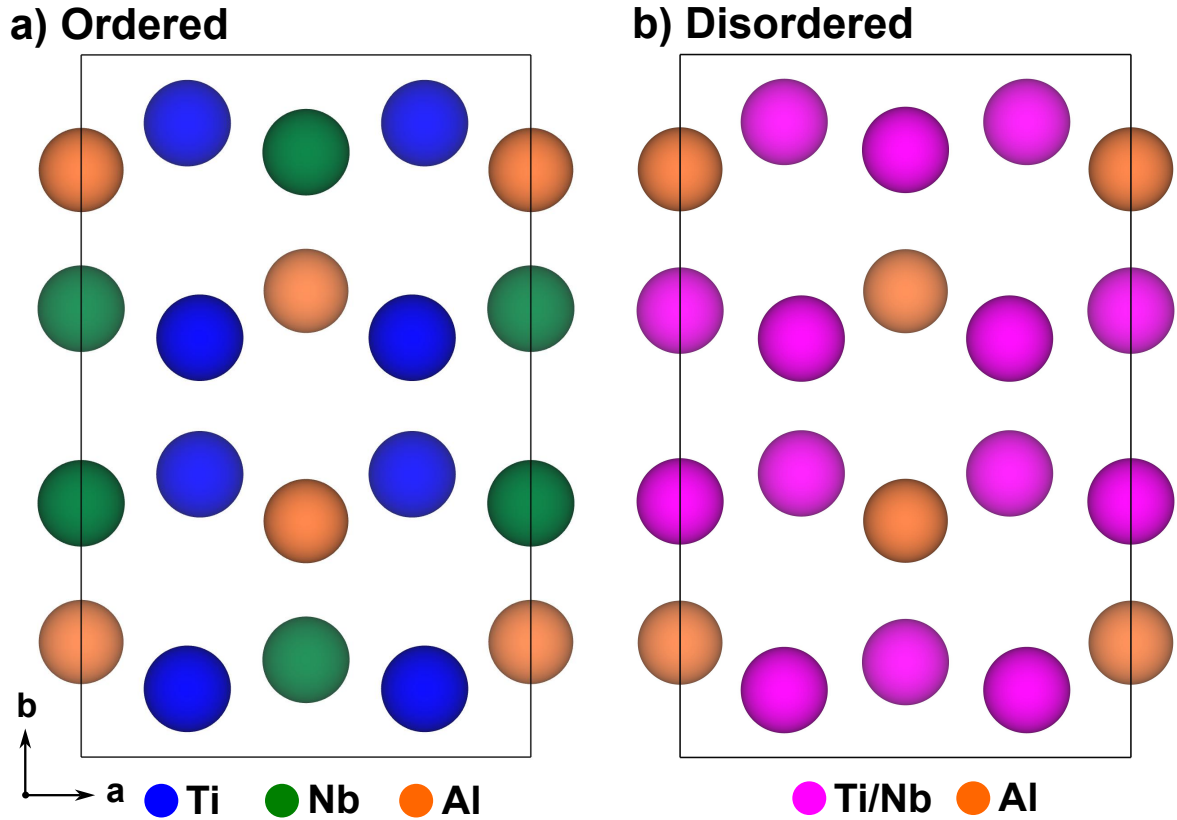

**Figure 1**: Structures of the ordered (a) and disordered (b) orthorhombic Ti$_2$AlNb intermetallic (O-phase). In the ordered state, the green and blue circles represent Ti and Nb atoms occupying the α (8g) and β (4c$_2$) sub-lattices, respectively. In the disordered state, purple circles represent Ti and Nb randomly occupying the α and β sub-lattices. In both phases, orange circles represent Al atoms occupying the γ (4c$_1$) sub-lattice.

The formation of Ti$_{Nb}$ and Nb$_{Ti}$ anti-sites plays a critical role in facilitating disorder within the O-phase. The extent of disorder significantly influences atomic mobility and, consequently, the creep resistance of the material. Note that creep is often enabled by diffusion, which requires atomic migration (or mobility) at the atomic scale – higher the atomic migration, lower the resistance to creep. Hence, a thorough understanding of the ordering behaviour of the O-phase, and the consequent impact on the atomic mobility, is essential for assessing its properties as utilised in high-temperature jet engine applications. Therefore, developing thermodynamic and/or kinetic strategies that promote the ordered O-phase and enhance the creep resistance of the O-phase based alloys is important in ensuring the practical use of Ti-based alloys.

In the present work, we explore the effect of various alloying elements on the ordering behavior of the O-phase and the consequent effects on atomic migration using density functional theory (DFT) and nudged elastic band (NEB) calculations. We correlate the energy associated with the $Ti_{Nb}+Nb_{Ti}$ anti-site defect formation (thermodynamics) and migration barrier of Ti/Nb (kinetics) in the presence of alloying elements to the eventual tendency of the O-phase to form the ordered structure. Note that the higher the values of anti-site formation energies and migration barriers, the higher the chances of suppressing the disordering on the Ti and Nb sub-lattices. We select the elements V, Cr, Fe, Mo, Ta, and W as alloying additions based on available experimental reports on their influence on the creep and tensile properties of O-phase alloys, their high melting points, and their intrinsic body-centred-cubic (BCC) crystal structure.[10-13] Specifically, we calculate the defect formation energy of $Ti_{Nb} + Nb_{Ti}$ anti-site pairs and the activation barrier for Ti/Nb migration in the presence of the alloying element. Notably, we observe that any alloying addition, when only considering the thermodynamics of anti-site formation, do not suppress the disorder on the Ti and Nb sub-lattices. However, when considering the kinetics of atomic diffusion, we find that Mo and W additions can suppress the disorder by reducing the mobility of Ti (by ≈4× than the pristine O-phase at an operating temperature of 823 K) and Nb (by ≈8×), thereby potentially improving the creep properties of O-phase based alloys.

## 2 Methods

We calculated the formation energy of the anti-sites considered using density functional theory (DFT[14,15]), as implemented in the Vienna ab initio simulation package (VASP[16,17]), and employing the projector-augmented-wave potentials.[18] We used an energy cut-off of 520 eV on the plane-wave basis set for all calculations and the Γ-centred Monkhorst-Pack scheme to sample the irreducible Brillouin zones with *k*-point meshes having a density of at least 48 points per Å (i.e., a minimum of 48 sub-divisions along a unit reciprocal lattice vector). We relaxed the cell volume, cell shape, and atomic positions for all bulk input structures until the atomic forces and total energies converged below |0.01| eV/Å and $10^{-5}$ eV, respectively. The electronic exchange-correlation interactions were treated with the Perdew-Burke-Ernzerhof generalized gradient approximation (GGA)[19] functional for all calculations. We did not preserve symmetry during any of the calculations and initialized all metal atoms to be in their high-spin ferromagnetic state.

The formation energy of any defect $E_f$ (in eV) is given by **Equation 1**, where $E_{defect}$ and $E_{bulk}$ are the total energies of a supercell with and without defect, respectively. $n_i$ represents the number of atoms of the $i$-species added (>0) or removed (< 0) to form the defect, with $\mu_i$ the corresponding chemical potential of the $i$-species. The $\mu_i$ considered in this work are referenced to GGA-calculated energies of the pure elements in their corresponding ground state structures at 0 K. For all defective structures, we relaxed only the atomic positions while keeping the lattice parameters of the relaxed bulk structure.

$$E_f = E_{defect} - E_{bulk} - \sum \mu_i n_i \qquad (1)$$

We performed the migration barrier calculations with the DFT-based NEB[20,21] method. Note that the activation barrier for diffusion, $E_d$, is related to the diffusivity ($D$) via the Arrhenius relation, i.e., $D = D_o \exp\left(-\frac{E_d}{k_B T}\right)$, where $D_o$, $k_B$, and $T$ are the diffusivity pre-factor, Boltzmann constant, and temperature, respectively. In a vacancy-mediated mechanism, $E_d$ is the sum of two terms, $E_v$ and $E_m$, which are the vacancy formation energy and the migration barrier, respectively, with NEB calculations yielding $E_m$. We considered seven images between the endpoints with a spring force constant of 5 eV/Å$^2$ between images. We used 2×1×2 supercells (64 atoms) of Ti$_2$AlNb in our NEB and defect calculations to ensure a minimum of ~8 Å distance between periodic images, thus reducing fictitious image-image interactions. We used a *k*-point density of 48 points per Å for relaxing the endpoints and 32 points per Å for the actual NEB calculation. We modelled the migration of both Ti and Nb atoms using a vacancy-mediated mechanism.

# 3 Results and Discussion

## 3.1 Ti$_2$AlNb structure

The initial Ti$_2$AlNb structure, as available in the inorganic crystal structure database (ICSD[22]), is disordered (**Figure 1**b). Hence, we utilized the OrderDisorderedStructureTransformation class from the pymatgen[23] package to enumerate the symmetrically distinct arrangements of Ti and Nb in the Ti$_2$AlNb structure. Upon enumeration, we obtained five symmetrically distinct orderings for the Ti$_2$AlNb structure, where four configurations can be considered to be partially disordered (i.e., at least one Nb atom occupies the 8g site) while the fifth configuration is fully ordered (i.e., all Nb atoms occupying 4c$_2$ sites, **Figure 1**a). **Table 1** presents the formation energies for all the different configurations, calculated with respect to the elemental references of Ti, Nb, and Al. Importantly, we find that the fully ordered configuration exhibits the lowest formation energy, indicating that it is the ground state (or the most stable) configuration. Additionally, our calculated formation energy is comparable to the value reported in the literature, further validating our calculations.[6]

**Table 1**: Formation energies of symmetrically distinct Ti-Nb configurations within the Ti$_2$AlNb structure obtained upon enumeration.

| Structure | Ordering | Formation energy (eV/atom) | Formation energy (literature) (eV/atom) |
|---|---|---|---|
| **Ti$_2$AlNb** | Partially disordered | -0.2669 | - |
| | | -0.2669 | - |
| | | -0.2677 | - |
| | | -0.2696 | - |
| | Fully ordered | -0.2823 | -0.2724 |

## 3.2 Anti-site defects

### 3.2.1 Defect: Nb$_{Ti}$ + Ti$_{Nb}$

The anti-site pair defect, Ti$_{Nb}$ + Nb$_{Ti}$, corresponds to the exchange of Nb and Ti atoms in the α and β sublattices and is emblematic of the disorder in the Ti$_2$AlNb structure. Given that the formation energy of a defect within a given structure reflects the energy penalty associated with the formation of that defect, a negative (or low positive) formation energy indicates a spontaneous formation of the corresponding defect in the structure. **Figure 2** plots the $Ti_{Nb} + Nb_{Ti}$ defect formation energy considering the presence of the alloy addition, Y (Y = V, Cr, Fe, Mo, Ta, and W), either in the Nb site (purple circle in panel a) or in the Ti site (panel c). The black arrows represent the exchange of the Ti and Nb atoms to form the $Ti_{Nb} + Nb_{Ti}$ defect. Note that the formation energy for the $Ti_{Nb} + Nb_{Ti}$ defect in pristine O-phase (i.e., without alloy addition) is ~249 meV, as indicated by the dashed black lines in panels b and d of **Figure 2**.

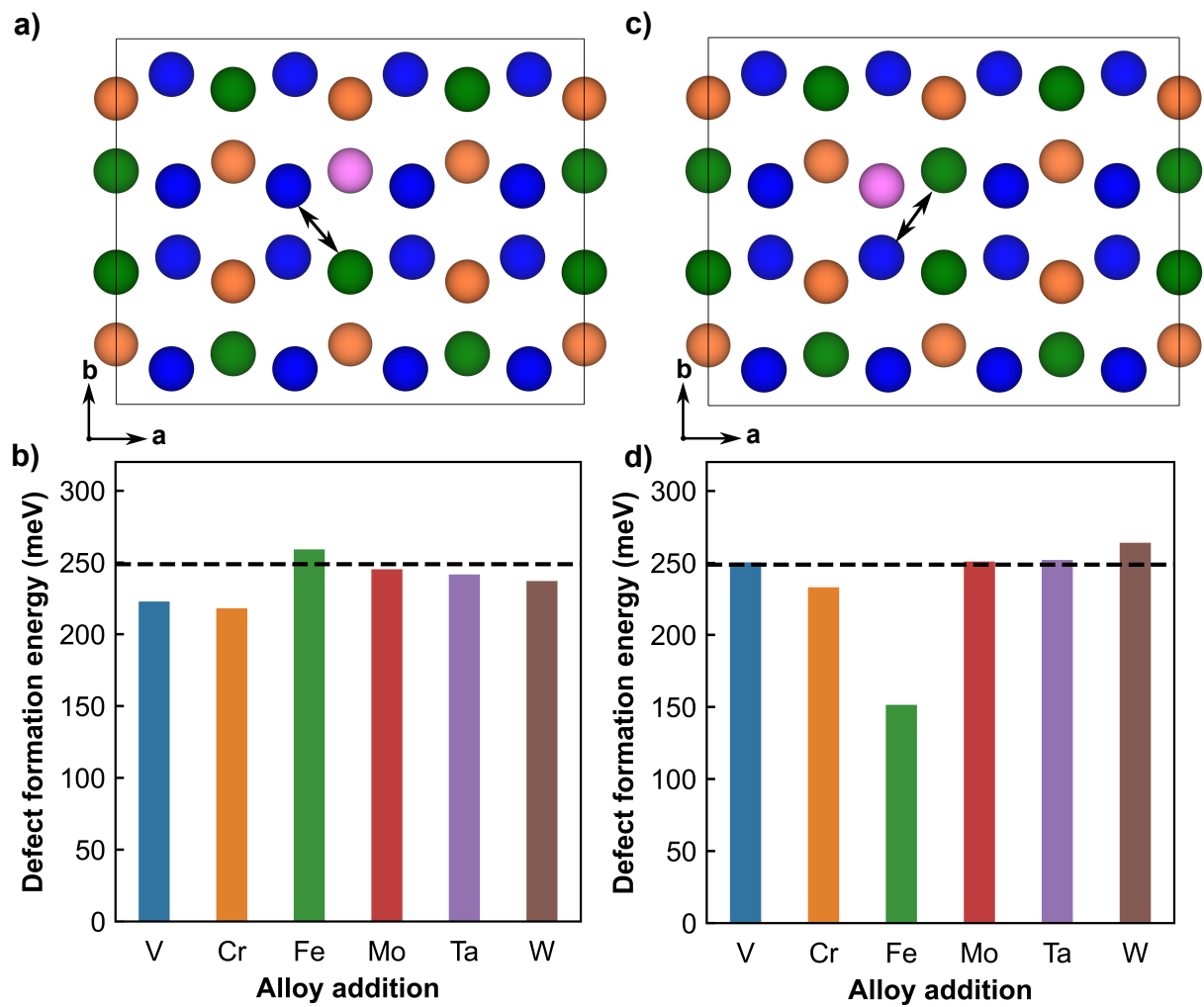

**Figure 2**: Schematic of $Nb_{Ti}+Ti_{Nb}$ defect formation and the corresponding formation energy when the alloy element (purple circle) is in Nb ($4c_2$) site (panels a and b) and Ti ($8g$) site (panels b and d). Black arrows display the exchange of Ti and Nb atoms causing disorder in the O-phase. The dashed black line in panels b and d indicates the $Ti_{Nb}+Nb_{Ti}$ defect formation energy in the pristine O-phase. Blue, green, and orange circles represent the Ti, Nb, and Al atoms, respectively.

To obtain the influence of the alloy element addition on the $Ti_{Nb}+Nb_{Ti}$ formation, we consider the exchange of Ti and Nb atoms that are nearest neighbours to the site occupied by the alloy element. Importantly, we do not find any of the alloying additions to increase the $Nb_{Ti}+Ti_{Nb}$ formation energy compared to the pristine, with Fe (on Nb, **Figure 2**b) and W (on Ti, **Figure 2**d) showing marginally higher formation energies. Moreover, several alloying additions facilitate the formation of $Nb_{Ti}+Ti_{Nb}$ defect compared to the pristine $Ti_2AlNb$ structure, such as V, Cr, Mo, Ta and W on the Nb site, and Cr and Fe on the Ti site. Thus, we do not expect any of the alloying additions considered, either in the Nb or Ti site, to thermodynamically suppress the $Nb_{Ti}+Ti_{Nb}$ defect effectively in the $Ti_2AlNb$ structure, signifying their lack of thermodynamic role in the eventual order-disorder transition of the O-phase and its creep resistance.

### 3.2.2 Defect: $Ti_{Nb} + Y_{Ti}$ and $Nb_{Ti} + Y_{Nb}$

Apart from reducing the formation of $Ti_{Nb} + Nb_{Ti}$ anti-site defects, the alloying additions should preferably not form anti-site defects themselves with the nearest Ti/Nb atoms, such as $Ti_{Nb}+Y_{Ti}$ or $Nb_{Ti}+Y_{Nb}$, given that the tendency of the O-phase to disorder will increase compared to the pristine O-phase if the alloying additions also form anti-site defects by themselves. In order to assess the tendency of alloy atoms to form anti-sites with neighbouring Ti/Nb sites, we evaluate the formation of a $Ti_{Nb} + Y_{Ti}$ defect, corresponding to the exchange of Y on the Nb site with a neighbouring Ti (panels a and b in **Figure 3**), and the formation of a $Nb_{Ti}+Y_{Nb}$ defect, signifying the exchange of Y on a Ti site with a neighbouring Nb (panels c and d). Dashed black lines in **Figure 3** indicate the formation energy of the $Nb_{Ti}+Ti_{Nb}$ defect in the pristine O-phase.

Our data (**Figure 3**b) reveal that alloying elements such as Mo and W exhibit higher defect formation energies, ~428 and ~393 meV, respectively, when forming anti-sites by exchanging with a neighbouring Ti, indicating that Mo and W are less prone to anti-site formation compared Ti-Nb exchange in pristine O-phase. Conversely, other alloying elements considered including V, Cr, Fe, and Ta form anti-sites by exchanging with a nearby Ti more readily than a Nb atom would in the pristine O-phase, signifying their higher propensity towards driving defect formation and disordering. In the case of alloy addition on the Ti site (**Figure 3**d), we observe all alloying elements to more spontaneously exchange a neighbouring Nb atom compared to a Ti atom exchanging Nb in the pristine O-phase, indicating a lack of thermodynamic suppression of anti-site defects by the alloying additions. Note that negative defect formation energies in **Figure 3**d do not indicate thermodynamic instability of the bulk structure, the negative values rather reflect the preference of the alloying atom to occupy a Nb site instead of a Ti site (see following sub-section) in the O-phase.

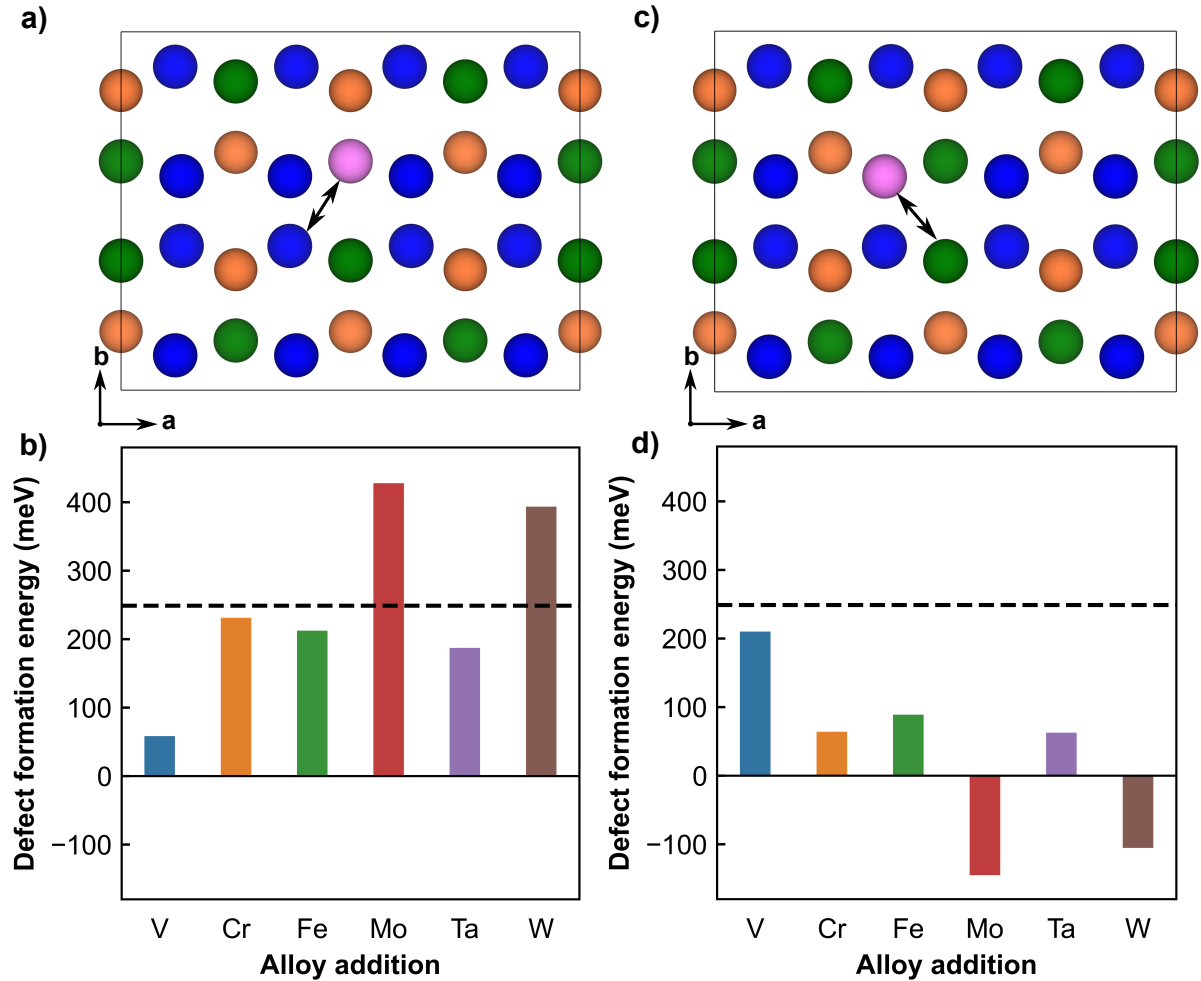

**Figure 3**: Schematic and defect formation energy of Ti$_{Nb}$ + Y$_{Ti}$ (panels a and b), and Nb$_{Ti}$+Y$_{Nb}$ (panels c and d). The dashed black line in panels b and d indicates the Ti$_{Nb}$+Nb$_{Ti}$ defect formation energy in pristine O-phase. Notations used in the figure are similar to **Figure 2**.

### 3.2.3 Site preference of alloying elements

In order to determine the preference of alloy additions for occupying either the Nb or the Ti site in the O-phase structure, we calculate the substitutional defect formation energy ($E_f$) for the elements considered (see **Equation 1**). $E_f$ represents the stability of the alloy atom when it replaces a host atom within the structure, with $E_{defect}$ in **Equation 1** representing the Ti$_2$AlNb structure with an alloying element placed in either a Ti or a Nb site. A lower $E_f$ value thus indicates a higher stability of the alloy addition in the given site, with the calculated values plotted in **Figure 4** for alloying elements placed in Nb (orange bars) and Ti (hashed green) sites.

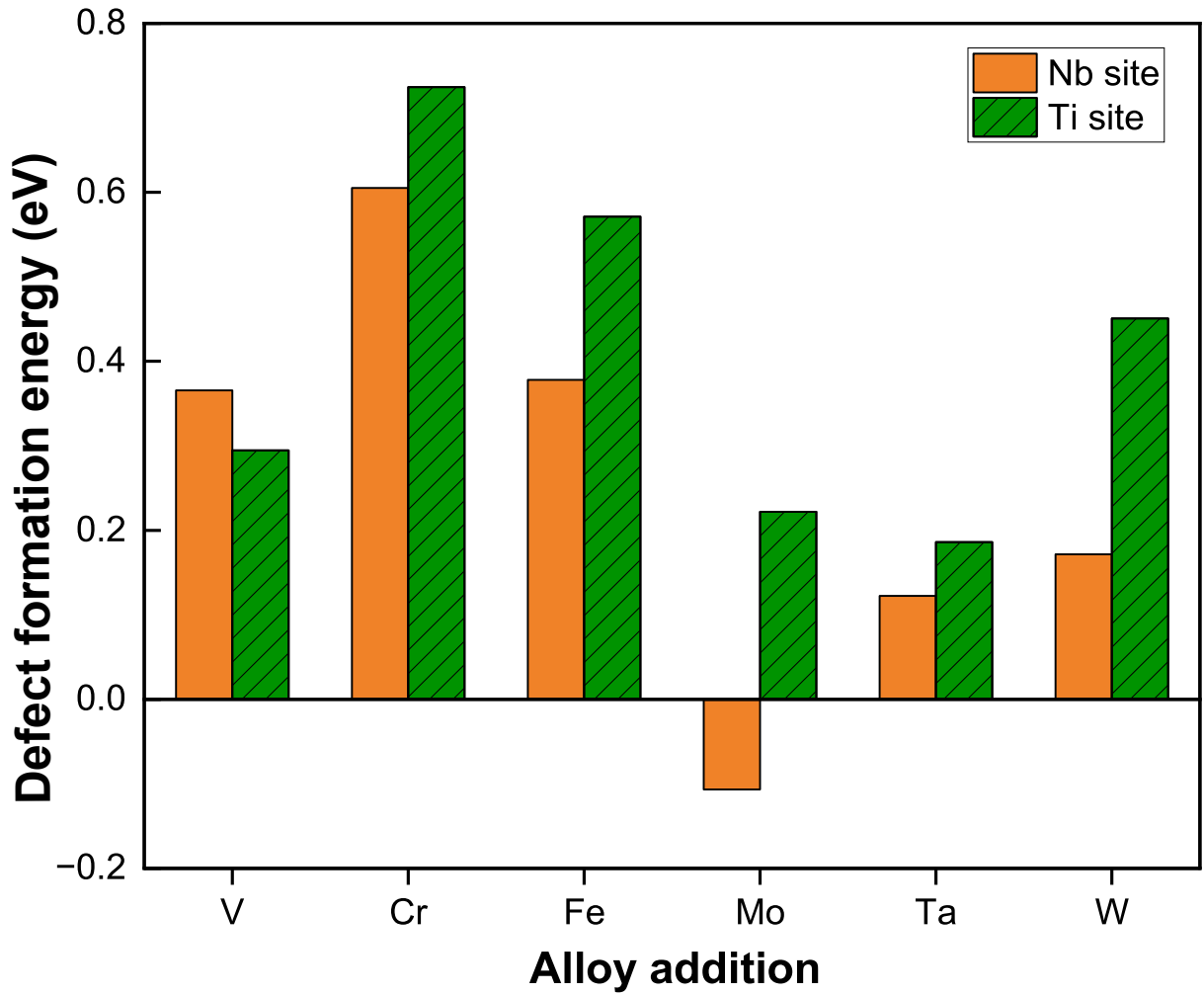

**Figure 4**: Substitutional defect formation energies of alloy atoms occupying the Nb (orange bars) or Ti (hashed green bars) site in the O-phase.

Importantly, we find that except for V, all other alloying additions prefer the Nb site compared to the Ti site, as indicated by the lower $E_f$ values for Nb site occupation in **Figure 4**, in agreement with our observation of lower/negative formation energies for the Nb$_{Ti}$+Y$_{Nb}$ defect (**Figure 3**d). Note that the negative $E_f$ for Mo occupying Nb site indicates that any Mo uptake by the O-phase will be a thermodynamically spontaneous process. So far, our findings indicate limited (or no) effectiveness of alloying additions thermodynamically reducing disorder in the O-phase, by considering various defect formation energies. Nevertheless, disorder can also be reduced kinetically, if there are significant migration barriers associated with the anti-site defect formation, which is the focus of the following sections.

### 3.3 Migration barriers

Given that Ti$_{Nb}$ or Nb$_{Ti}$ anti-sites are required for increasing the disorder of the O-phase, a significant kinetic barrier for the migration of Ti (Nb) to a neighbouring Nb (Ti) site, especially in the presence of a suitable Y atom, can reduce the extent of anti-site formation and disorder in the O-phase. Thus, we calculate the barriers for Ti migrating to a nearby vacant Nb site and vice-versa, in the presence/absence of alloying elements in adjacent sites. Note that we are assuming all migrations to occur via an isolated vacancy, while more complex mechanisms involving multiple vacancies are possible but are computationally expensive to model.

Panels a and b of **Figure 5** illustrate the migration pathway (band of cyan circles) and calculated migration barriers ($E_m$, bar charts) for Ti migration to the nearest vacant Nb site, while panels c and d highlight the pathway and barriers for Nb migration to the nearest vacant Ti site. The maroon circle in panels a and c of **Figure 5** signify the alloy atom occupying a Nb site, since most alloying additions considered in this work prefer the Nb site (**Figure 4**). Black arrows in **Figure 5** indicate the direction of Ti/Nb migration. Dashed black lines in panels b and d of **Figure 5** indicate the $E_m$ for Ti hopping to the nearest vacant Nb site (~888 meV) and for Nb hopping to the nearest Ti site (~780 meV), respectively, in the pristine O-phase. Thus, alloying additions resulting in $E_m$ that are larger than the pristine O-phase can potentially reduce disorder in the structure.

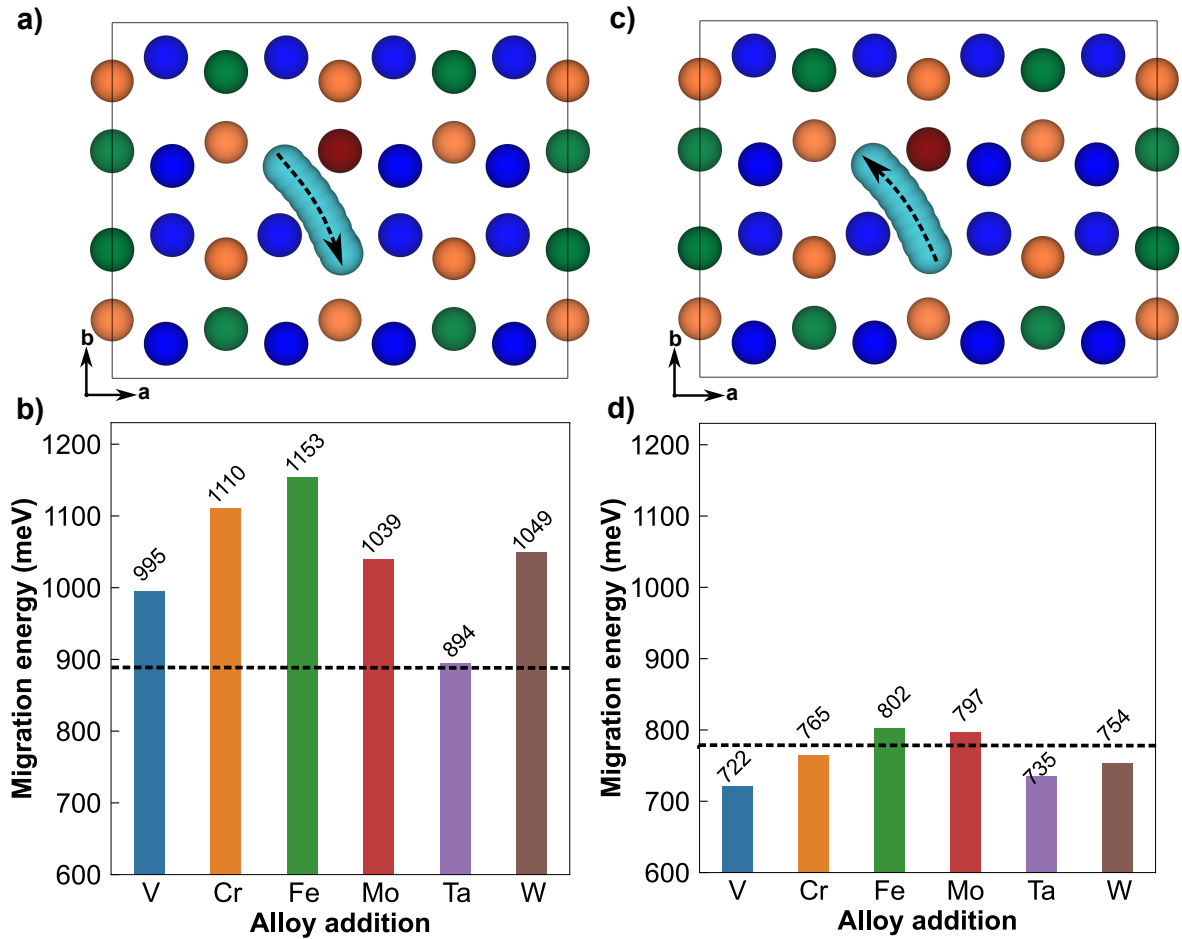

**Figure 5**: Calculated migration pathway and barriers for Ti hopping to a nearest vacant Nb site (panels a and b) and Nb hopping to a nearest vacant Ti site (panels c and d) in the O-phase. Maroon circles are alloy atoms and dashed black lines are calculated barriers in the pristine O-phase.

In comparison with the pristine O-phase, all alloying additions considered, except Ta, cause an increase in the migration barrier for Ti hopping to the nearest vacant Nb site, indicating that these alloying elements can potentially suppress disorder in the O-phase kinetically. For example, addition of V, Cr, Fe, W, and Mo cause an increase in $E_m$ by ~107, ~222, ~265, ~151, and ~161 meV, respectively (**Figure 5**b), compared to the pristine O-phase, which can correspond to at least 2 orders of magnitude lower Ti migration at room temperature (as per the Arrhenius relation assuming constant $E_v$). On the other hand, we do not observe any kinetic suppression of Nb migration to an adjacent vacant Ti site due to the presence of alloying elements, suggesting that Nb atoms will form Nb$_{Ti}$ anti-site defects as long as there are vacant Ti sites that are neighbouring them. Indeed, V, Cr, Ta, and W addition cause a drop in $E_m$ by ~58, ~15, ~45, and ~26 meV, respectively (**Figure 5**d), compared to the pristine O-phase, highlighting an enhancement of Nb hopping by at most an order of magnitude at room temperature. Note that Fe and Mo marginally suppress Nb migration by ~22 and ~17 meV, respectively, compared to the pristine O-phase, thus making them the only alloying additions considered that can potentially hamper both Ti and Nb migration. With respect to V addition, we placed the V atom on a Nb site although it prefers the Ti site (**Figure 4**). However, we expect a V placement on a Ti site to only create marginal changes in the $E_m$ calculated given the similarity of both the substitutional defect formation energy of V in both the Nb and Ti sites (**Figure 4**).

To further understand the increase in $E_m$ associated with alloying additions in the case of Ti hopping, we calculate the binding energies of the alloy atom that is present in the Nb site with an adjacent Ti (hashed purple bars) or Nb (blue) vacancy and plot them in **Figure 6**. Note that binding energies represent the energy difference between entities (e.g., alloy atom and Ti/Nb vacancy) that occupy 'adjacent' sites in a solid lattice and their isolated existence in the same lattice. Thus larger (i.e., more negative) binding energies typically indicate entities that are strongly bound. Interestingly, all alloying additions exhibit negative binding energies (i.e., are strongly bound, **Figure 6**) with an adjacent Nb vacancy, indicating the difficulty of Nb vacancy to be displaced away from the alloy atom, which correlates with the enhanced barriers associated with Ti atom hopping into a vacant Nb site (thus displacing the Nb vacancy, **Figure 5**b). On the other hand, all alloying additions, except Fe, exhibit positive binding energies (i.e., weakly bound) with an adjacent Ti vacancy, which also correlates with the reduced barriers associated with Nb atom hopping into a vacant Ti site (**Figure 5**d). Thus, any kinetic suppression of disorder in the O-phase is likely associated with vacancy trapping that reduces the extent of anti-site defect formation by an added alloy element.

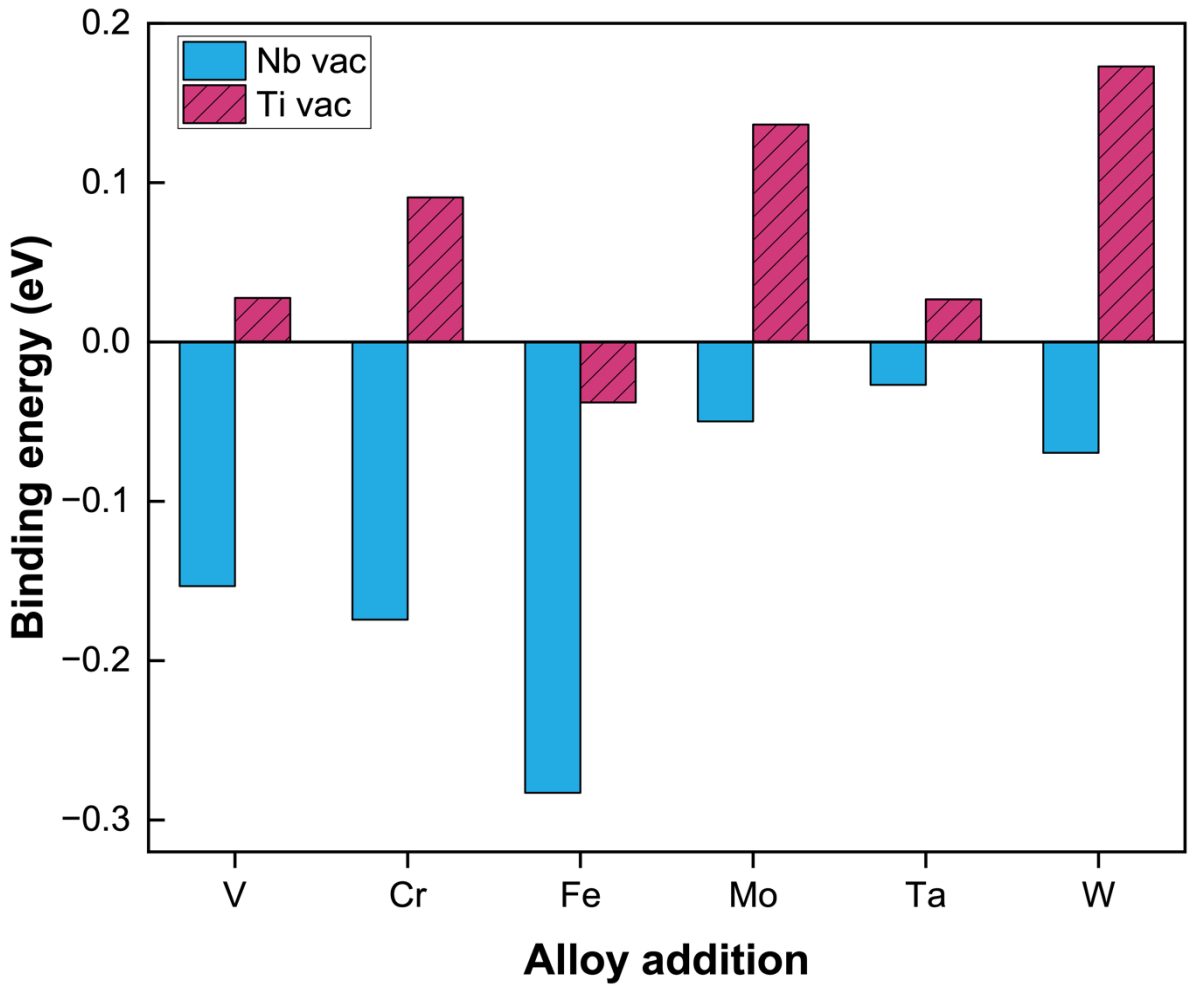

**Figure 6**: Binding energies of alloy atom and Nb/Ti vacancy (vac) pairs in the O-phase. Blue and hashed purple bars correspond to Nb and Ti vac. The alloy atom is in the Nb site.

### 3.4 Overall activation barrier

Given that vacancy-based activation barriers for diffusion in solids is the sum of the $E_v$ and $E_m$ terms, we estimate the $E_d$ in the presence of various alloying elements occupying the Nb site and plot them for Ti (panel a) and Nb (panel b) diffusion in **Figure 7**. Note that for Ti (Nb) diffusion, a Nb (Ti) vacancy is required. Thus, $E_d$ for Ti (Nb) diffusion (green bars in **Figure 7**) is the sum of $E_v$ for Nb (Ti) vacancy formation (red bars) and the $E_m$ for Ti (Nb) migration (blue bars). Dashed horizontal lines in **Figure 7** indicate $E_v$ (red colour), $E_m$ (blue), and $E_d$ (green) in the pristine O-phase for Ti (panel a) and Nb (panel b). $E_m$ values plotted in **Figure 7** are identical to **Figure 5**. In the absence of alloying additions, we calculate the overall $E_d$ for Ti and Nb diffusion in the pristine O-phase to be 3.09 and 3.42 eV, respectively, consistent with the range of values (3.08-3.58 eV) reported in literature for alloys based on the O-phase.[10-13]

Additionally, the Monkman-Grant relation[24] relates the steady state creep rate ($\dot{\epsilon}$ in s$^{-1}$) and the time to fracture ($t_f$ in s) of a system to a constant ($C_{MG}$) as $\dot{\epsilon}\, t_f = C_{GM}$, assuming a constant strain to failure. Note that $\dot{\epsilon}$ roughly scales in an Arrhenius manner, as $\dot{\epsilon} \propto \exp\left(-\frac{E_C}{k_B T}\right)$, where $E_C$ is the activation energy associated with creep. In systems dominated by diffusion creep, $E_C \approx E_d$. In order to achieve a tenfold increase in fracture time, $\dot{\epsilon}$ needs to reduce by tenfold, which necessitates an increase in $E_d$ (or $E_C$) by 173 meV at an operating temperature of 823 K. Thus, alloying additions that can cause an effective increase in $E_d$ by 173 meV (or similar) are desirable. Note that the Monkman-Grant relation is an approximation used for estimating creep rates or rupture times and is useful in providing qualitative guidance.

Our calculated data in **Figure 7** indicates that $E_v$ is the dominant contributor to $E_d$ than $E_m$. For example, Fe exhibits significantly higher $E_m$ for Ti hopping (**Figure 7**a) but the overall $E_d$ for Ti is lower than the pristine O-phase since Nb vacancy formation is easier near Fe resulting in a lower $E_v$. While V and Ta exhibit a similar trend to Fe for Ti hopping, Cr, Mo, and W exhibit larger $E_d$ than the pristine O-phase. Notably, both Mo and W addition increase $E_d$ by ~100 meV (specifically by 101 and 92 meV with Mo and W addition, respectively) compared to the pristine O-phase signifying a ≈4× reduction in the diffusivity at 823 K. Also, Mo and W both exhibit higher (lower) $E_m$ ($E_v$) than the pristine O-phase. Thus, if Ti diffusivity is the key bottleneck determining $\dot{\epsilon}$, adding Mo and W will likely cause a ≈4× reduction in the observed $\dot{\epsilon}$.

In the case of Nb diffusion (**Figure 7**b), we observe Cr, Mo, and W (V, Fe, and Ta) additions to exhibit $E_d$ higher (lower) than the pristine O-phase, with $E_v$ being the dominating contributor to $E_d$, similar to Ti diffusion. Addition of Cr, Mo, and W causes the $E_v$ to be higher than the pristine O-phase while the $E_m$ values are similar. Importantly, Cr, Mo, and W cause an increase in $E_d$ of ~75, ~153, and ~147 meV, respectively, corresponding to a ≈3×, 8×, and 8× drop in Nb diffusivity at 823 K. Thus, if Nb diffusivity is the key bottleneck for creep, the associated $\dot{\epsilon}$ will exhibit ≈3×, 8×, and 8× drop with the addition of Cr, Mo, and W. Finally, among the alloying additions considered, given that Mo and W reduce both Ti and Nb diffusivity, we expect Mo and W addition to Ti2AlNb-based O-phase alloys to be a promising pathways to kinetically reduce creep and improve the performance of O-phase alloys at high temperatures.

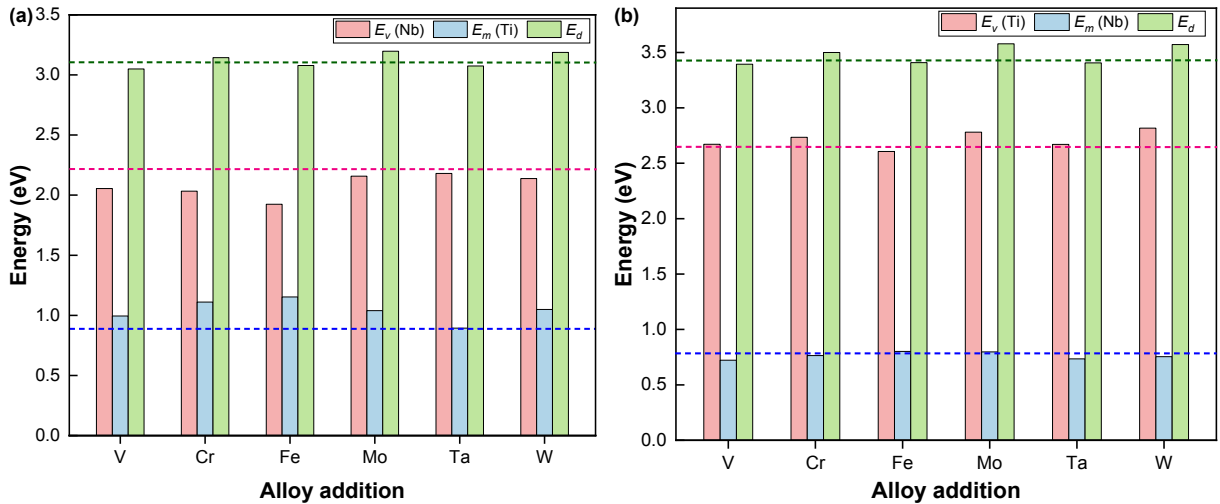

**Figure 7**: Vacancy formation energy ($E_v$), migration barrier ($E_m$), and the overall activation energy for diffusion ($E_d$) associated with Ti (panel a) and Nb (panel b) diffusion with various alloying additions in O-phase. Blue, red, and green dashed lines indicate $E_m$, $E_v$, and $E_d$, respectively, in pristine O-phase. The alloy atom occupies a Nb site in all calculations.

## 4 Conclusion

Alloys based on the orthorhombic Ti2AlNb intermetallic phase are potential materials for aerospace applications that can substitute a part of Ni-based superalloys in jet engines. However, O-phase alloys are prone to disordering at high temperatures (>773 K) due to the formation of Ti$_{Nb}$ and Nb$_{Ti}$ anti-site defects.[5] The lattice disorder, typically confined to the Ti and Nb sublattices in the Ti2AlNb structure, reduces the creep resistance of O-phase containing alloys. Hence, it is vital to identify strategies to suppress disorder in the O-phase, which is the focus of this work.[8]

Specifically, we investigated the effect of alloying additions, namely, V, Cr, Fe, Mo, Ta, and W, on the thermodynamics of anti-site formation and the kinetics of Ti and Nb diffusion in the Ti2AlNb structure using density functional theory calculations. Upon verifying that the fully ordered structure was indeed the ground state of Ti2AlNb, we examined the tendency of the alloying element to facilitate the formation of both Ti$_{Nb}$+Nb$_{Ti}$ pairs as well as Ti$_{Nb}$+Y$_{Ti}$ and Nb$_{Ti}$+Y$_{Nb}$ pairs. Our results indicated that none of the alloying additions considered can suppress anti-site formation and hence can't reduce disorder thermodynamically. Additionally, we observed the tendency of all alloying elements (except V) to occupy the Nb site in the $Ti_2AlNb$ structure. Importantly, we found that Mo and W addition can kinetically suppress disorder and improve the creep resistance of the O-phase, by reducing the Ti (by ≈4× at 823 K) and Nb (by ≈8×) diffusivities compared to the pristine O-phase, thus providing a promising strategy to engineer creep-resistant Ti-based alloys.

## Data and code availability

All data and codes associated with this work will be made available upon a reasonable request to the corresponding author.

## Conflict of interest

The authors declare no conflicts of interest relevant to this work.

## Acknowledgments

G. S. G. acknowledges financial support from the Defence Research and Development Organization, Government of India, under sanction number DFTM/02/3125/M/08/HTM-04. A. C. thanks the Ministry of Human Resource Development, Government of India, for financial assistance. The authors acknowledge Prof. Satyam Suwas and Dr. Surendra Kumar Makineni in the Department of Materials Engineering, Indian Institute of Science (IISc) for fruitful discussions. The authors acknowledge the computational resources provided by the Supercomputer Education and Research Centre, IISc, for enabling some of the density functional theory calculations showcased in this work. A portion of the calculations in this work used computational resources of the supercomputer Fugaku provided by RIKEN through the HPCI System Research Project (Project ID hp220393 and hp240314).